\documentclass{article}

\usepackage[english]{babel}

\usepackage[a4paper,top=2cm,bottom=2cm,left=3cm,right=3cm,marginparwidth=1.75cm]{geometry}

\usepackage{amsmath}
\usepackage{amssymb}
\usepackage{graphicx}
\usepackage{booktabs}
\usepackage{authblk}
\usepackage{lineno}
\usepackage[numbers]{natbib}
\usepackage{tabto}
\usepackage[colorlinks=true, allcolors=blue]{hyperref}

\begin{document}
\title{Knowledge Transfer, Knowledge Gaps, and Knowledge Silos in Citation Networks}
\author[1,2,*]{Eoghan Cunningham}
\author[1,2]{Derek Greene}


\affil[1]{School of Computer Science, University College Dublin, Dublin, Ireland}

\affil[2]{Insight Centre for Data Analytics, Dublin, Ireland}

\affil[*]{eoghan.cunningham@insight-centre.org}
\maketitle

\begin{abstract}
The advancement of science relies on the exchange of ideas across disciplines and the integration of diverse knowledge domains. However, tracking knowledge flows and interdisciplinary integration in rapidly evolving, multidisciplinary fields remains a significant challenge. This work introduces a novel network analysis framework to study the dynamics of knowledge transfer directly from citation data. By applying dynamic community detection to cumulative, time-evolving citation networks, we can identify research areas as groups of papers sharing knowledge sources and outputs. Our analysis characterises the life-cycles and knowledge transfer patterns of these dynamic communities over time. We demonstrate our approach through a case study of eXplainable Artificial Intelligence (XAI) research, an emerging interdisciplinary field at the intersection of machine learning, statistics, 
and psychology. Key findings include: (i) knowledge transfer between these important foundational topics and the contemporary topics in XAI research is limited, and the extent of knowledge transfer varies across different contemporary research topics; (ii) certain application domains exist as isolated "knowledge silos"; (iii) significant "knowledge gaps" are identified between related XAI research areas, suggesting opportunities for cross-pollination and improved knowledge integration. By mapping interdisciplinary integration and bridging knowledge gaps, this work can inform strategies to synthesise ideas from disparate sources and drive innovation. More broadly, our proposed framework enables new insights into the evolution of knowledge ecosystems directly from citation data, with applications spanning literature review, research planning, and science policy.
\end{abstract}


\section{Introduction}
\label{sec:intro}

The advancement of science is driven by the exchange of ideas across disciplines and the integration of diverse knowledge domains \cite{shi2023surprising}. Understanding the evolution of research fields and the transfer of knowledge between them is crucial for effective interdisciplinary collaboration and scientific progress. Interdisciplinary research, which integrates methods and expertise from different domains, is highly valued for its potential to drive innovation and impact \cite{lariviere2015long}. However, tracking the development of rapidly evolving, multidisciplinary fields poses significant challenges, as static taxonomies often fail to capture the dynamic nature of these research areas \cite{leydesdorff2007visualization}, while traditional content-based topic modelling methods ignore the knowledge transfer encoded by citation relations.

In this work, we develop novel methods to analyse knowledge transfer in rapidly evolving, highly interdisciplinary research fields. Our approach aims to uncover knowledge gaps and knowledge silos \cite{nature2016silos}, where effective knowledge transfer is not taking place. Rather than relying on prescribed discipline taxonomies or traditional topic models, we leverage dynamic community detection techniques from social network analysis \cite{dakiche2019tracking} to identify and track the evolution of research topics directly from the citation network. Citation networks offer a unique perspective on knowledge transfer, as they represent the flow of information and ideas through the literature. By grouping papers into dynamic communities based on their shared knowledge sources and outputs, we can delimit \textit{research areas} as they naturally emerge and evolve over time. Such \textit{research areas}, as determined by community finding, differ from `\textit{research topics}', as determined by topic modelling or other natural language processing (NLP) techniques. For example, two distinct \textit{research areas} that exhibit limited knowledge transfer may pertain to the same \textit{research topic}. Accordingly, research paper text remains a necessary component of citation network analysis \cite{bruggeman2012detecting,vilhena2014finding}. In our work, we leverage research paper text content for the validation and interpretation of the dynamic communities. By analysing the interactions between these communities, we can gain insights into the nature of knowledge transfer in a body of research and how this process evolve over time. Specifically, we measure the extent to which contemporary topics build upon foundational research, we identify isolated knowledge silos, and uncover significant knowledge gaps between related research areas.

A key methodological contribution of this work is the new perspective we provide on dynamic community finding algorithms to facilitate their application to the unique context of mapping knowledge transfer in citation networks, which exhibit cumulative growth and content-rich nodes. The primary goal of many existing applications of dynamic community finding to citation networks is to benchmark the performance of some proposed community finding algorithm \cite{hopcroft2004tracking,cordeiro2016dynamic}, rather than to understand the real-world dynamics of interdisciplinary or multidisciplinary research interactions. Therefore, they frequently overlook the specific properties and nuances of cumulative citation networks and fail to explain how to interpret the resulting dynamic communities. Further, many instances of citation network analysis have been found to ignore the article content \cite{vilhena2014finding}. We propose dynamic community finding methods specific to the study of knowledge transfer in citation networks. In our methods we consider the unique temporal properties, and rich paper content inherent to cumulative citation networks to identify and characterise the life-cycles, content coherence, centrality, and other knowledge transfer patterns of dynamic communities (or research areas) over time.
We demonstrate the utility of our methods by investigating the following research questions relating to knowledge transfer in any rapidly evolving, interdisciplinary research field. 

\begin{enumerate}
    \item How do contemporary topics in the field rely on foundational research?
    \item What are the research areas that are most isolated in terms of knowledge transfer within the literature?
    \item Is there evidence of knowledge gaps between otherwise related research topics? 
\end{enumerate}

We choose eXplainable Artificial Intelligence (XAI) as a case study, as it represents a highly interdisciplinary research area that draws on concepts from a diverse set of foundational topics and that has important implications and applications across many fields of study \cite{samek2019towards}. In the context of this application, the three research questions above take the following forms.

(i)\textit{ How do contemporary topics in XAI rely on foundational research in psychology, statistics, and computer science?}
Despite the recent rapid growth of explainable artificial intelligence research, the field has its roots in topics such as psychology and cognitive science (the psychology of explanation), computer science and statistics (model interpretability), and political science and social science (ethics, governance and accountability). It is pertinent to understand the extent to which current XAI research leverages and builds on these studies. 

(ii)\textit{ What are the most isolated research areas in XAI in terms of knowledge transfer?}
The field of XAI is highly multidisciplinary -- methods and concepts from many research disciplines are represented. We aim to identify the most isolated research areas in the literature as potential "knowledge silos", which exhibit minimal knowledge transfer with other research areas (foundational or contemporary).

(iii)\textit{ What are the most significant knowledge gaps in the XAI literature?}
Given the accelerated rate of publication of XAI-related research, it may be challenging for authors to keep abreast of the research outputs that relate to their own. As such, short sighted reading and citation patterns can develop which could lead to knowledge gaps. We seek to identify knowledge gaps by modelling the probability of knowledge transfer between research areas according to their content similarity and citation neighbourhoods. Thus, pairs of research areas that exhibit substantially less knowledge transfer than predicted are concluded to have knowledge gaps. 


\subsection{Related Work}
\label{sec:background}

Mapping the structure and evolution of scientific fields has been an active area of research. Prescribed taxonomies and discipline classifications have been widely used to categorise papers into broader subjects or research areas \cite{chakrabort2013computer, cunningham2022author, taher2023embedding}. For example, Microsoft Academic and Web of Science maintain large, 
subject or `field-of-study' classifications for articles that are readily available and provided at multiple levels of detail. However, such top-down, static, taxonomies struggle to accurately capture the organic evolution of research topics \cite{ledeysdorff2008global, leydesdorff2006can, milojevic2019practical}, as disciplinary borders have been shown to be changing constantly \cite{rosvall2008maps}. Alternative, bottom-up methods have been developed to identify the constituent research topics in some larger corpus. In particular, keyword-based methods identify research topics as groups of commonly co-occurring keywords \cite{chae2016cannibalism, mane2004mapping, song2014analyzing}, while topic modelling techniques can be applied to research paper abstracts or full texts \cite{deligiannis2021visualising, vazquez2022validation}. Such bottom-up methods have advantages over static, prescribed classifications as they can be adapted to the specific application/dataset. Some works have developed these methods to study the evolution of topics and topic similarity over time \cite{song2014analyzing, taher2023embedding, wang2022matrixsim}. However, topic modelling methods come with challenges related to model selection and validation \cite{vazquez2022validation}. Moreover, they fail to recognise the transfer of knowledge inherent in a corpus of scientific research, in particular: the knowledge transfer represented by citation relationships.  

Network analysis approaches have emerged as powerful tools for mapping the landscape of scientific research directly from citation patterns \cite{chakraborty2013overcite,jung2013analyzing,quattrociocchi2012selection,asatani2018detecting}. Community detection algorithms aim to identify densely connected groups of nodes in networks, representing communities or clusters with strong internal connections and relatively fewer external connections \cite{fortunato2010community}. These algorithms have found numerous applications in various domains, including social network analysis \cite{bae2017scalable,fraisier2017uncovering}, biological networks \cite{sah2014exploring}, and network science \cite{fortunato2010community}. By modelling scientific literature as networks of papers connected by citations, community detection methods can uncover the underlying research areas present as groups of related papers that share knowledge sources and/or knowledge outputs. As research fields evolve over time, the underlying network structures and community memberships change, necessitating the study of dynamic communities. Dynamic community detection algorithms have been developed to track the evolution of communities in social networks \cite{dakiche2019tracking,palla2007quantifying}. Building on works in dynamic or `evolutionary' clustering \cite{chakrabarti2006evolutionary,spiliopoulou2006monic}, many dynamic community finding techniques partition the data into a series of temporal snapshots or time windows \cite{dakiche2019tracking}. By matching communities identified in adjacent networks snapshots, dynamic community life-cycles are identified as sequences of matched snapshot communities \cite{palla2007quantifying,greene2010tracking,quattrociocchi2012selection}. Existing literature has proposed various metrics specific to dynamic communities, such as `\textit{stability}'\cite{michel2023metrics}, `\textit{stationarity}'\cite{palla2007quantifying}, and `\textit{density}' \cite{michel2023metrics}, and developed methods for identifying and characterising life-cycle `events', such as community births, deaths, merges, and splits.

Several studies have applied dynamic community analysis techniques to citation networks. Many of the earliest examples include dynamic citation networks as case-studies, where the specific focus of the work is to demonstrate the performance of their proposed community finding methods \cite{hopcroft2004tracking,rosvall2008maps}. Subsequent works have had a more explicit focus on mapping or predicting changes in research networks. For example, Chakraborty et al.~\cite{chakrabort2013computer} measure changes in various citation network metrics for different fields-of-study, as prescribed by the Microsoft Academic Graph. Similar works track changes in community metrics for fields identified in a bottom-up manner using community detection \cite{quattrociocchi2012selection, tan2022tracking}, while others focus on predicting future changes \cite{jung2013analyzing}. Further examples of dynamic community finding in research networks build multi-partite graphs of papers, authors, concepts and/or venues, in order to uncover the social dynamics that define the formation of research areas \cite{shi2015weaving,chakraborty2013overcite}. Limited research exists to date that leverages dynamic community finding in citation networks to study knowledge transfer across disciplines. 

Effective knowledge transfer and integration across different disciplines are crucial for addressing complex scientific challenges and driving innovation \cite{lariviere2015long,shi2023surprising}. However, interdisciplinary research often faces significant hurdles, such as insular reading and citation practices \cite{leischow2008systems}. Such barriers can lead to the formation of knowledge gaps and silos, hindering scientific progress and productivity \cite{portenoy2022bursting, rodriguez2022speed}. Identifying and bridging these knowledge gaps is essential for fostering interdisciplinary collaboration and facilitating the cross-pollination of ideas and methods. Several studies have recognised the presence of knowledge silos and their detrimental effects on scientific research, emphasising the need for approaches to map and understand knowledge transfer dynamics \cite{leischow2008systems, portenoy2022bursting, rodriguez2022speed}. However, limited research exists to date that studies knowledge transfer from the perspective of dynamic citation network analysis.

In this work, we extend existing methods for finding dynamic communities to specifically map knowledge transfer and identify knowledge gaps in citation networks. Crucially, citation networks have unique characteristics, such as cumulative growth, where papers and citations are never removed from the network, and content-rich nodes, where papers include substantial textual information. These features necessitate novel approaches to community detection and tailored strategies for interpreting their outputs. In particular, the cumulative growth of citation networks must be acknowledged during the application of any existing community finding methods. Further, textual metadata available for papers has been highlighted as an important resource for citation network analysis \cite{bruggeman2012detecting,vilhena2014finding}. Leveraging this textual information can help validate and interpret the identified communities. Specifically, dynamic community metrics which consider the content of the papers in addition to community membership and network structure, could serve to bridge the gap between network-based methods of tracking research evolution and other NLP-based approaches (e.g.~\cite{song2014analyzing,taher2023embedding,wang2022matrixsim}).

\section{Methods and Materials}
\label{sec:methods}

Our framework combines dynamic community detection on citation networks with natural language processing of scholarly text. We model a \textit{cumulative citation network} as a discrete-time dynamic network, with time steps representing snapshots of the network state. Applying community detection to these time step networks identifies groups of papers representing distinct research areas based on shared citation patterns. Leveraging paper text allows us to validate, interpret, and characterise these communities. Tracking communities over time reveals the life-cycles of research areas, while analysing inter-community interactions exposes knowledge transfer patterns. 

The remainder of this section is structured as follows. Section \ref{sec:community_detection} describes the process of identifying research areas as communities at each time step. Sections~\ref{sec:labelling_communities} and \ref{sec:characterising_step_communities} outline the procedures for labelling and characterising these step communities respectively. In Section~\ref{sec:community_interaction_network}, we present our proposed method for tracking knowledge transfer between research areas, through the use of a \textit{community interaction network}. 
Section~\ref{sec:tracking_communities} explains tracking step communities to construct dynamic community life-cycles, while Section~\ref{sec:characterising_dynamic_communities} discusses characterising those dynamic communities. Subsequently, in Section~\ref{sec:research_questions} we discuss how the proposed methods can allow us to address the key research questions originally introduced in Section \ref{sec:intro}. Finally, in Section~\ref{sec:data} we discuss the construction of the dataset which is considered later in our case study in Section~\ref{sec:results}. The code and data used in our analysis are available at \cite{data_code}.

\subsection{Step Communities}
\label{sec:dyn_citation_graphs}
To investigate the life-cycles of research areas within a specific body of work, we require a citation network that evolves over time. In this work, we adopt a discrete time dynamic network such that the citation network is described by a sequence of time step graphs $\mathbb{G} = \{ \dots, G_{t-1}, G_t, G_{t+1}, \dots\}$, where each time step graph is defined by a set of nodes and edges $G_t = (V_t, E_t)$. Each set of nodes $V_t$ contains the papers present in the network at time step $t$ and the set of edges $E_t$ represent the citations between them. Due to the cumulative nature of citation networks, discrete time dynamic citation networks represent a unique type of dynamic network. Specifically, nodes (papers) and connections (citations) are never removed from the network. As such, the set of time step graphs are cumulative, i.e. $G_{t-2} \subseteq G_{t-1} \subseteq G_t$. In this application, we divide the citation network by year, such that the step graph denoted by $G_{2010}$ contains all papers in the dataset published before 2011.  

\subsubsection{Finding Step Communities}
\label{sec:community_detection}

To identify the research areas present in the literature at each time step, we apply community detection to the corresponding time step graph $G_t$ in the dynamic citation network. Specifically, we use the OSLOM algorithm \cite{lancichinetti2011finding} to extract overlapping communities that represent densely connected groups of papers sharing knowledge sources and outputs.
OSLOM is well-suited for our analysis due to several desirable properties. First, it can detect communities following a broad range of size distributions, avoiding the bias of some algorithms towards few large communities. Second, it identifies hierarchically nested communities, capturing the multiscale organisation common in citation networks. Third, it allows for overlapping communities, reflecting how papers can belong to multiple research areas. We refer to the communities discovered in time step graph $G_t$ as step communities $\mathbb{C}_t = \{C_1^t, C_2^t, .... C_n^t\}$

We initialise OSLOM for each time step using the communities identified in the previous time step. This leverages the temporal continuity expected in the evolution of research areas. To ensure consistent community identification across time steps, we use a fixed set of hyperparameters for OSLOM. This avoids having apparent dynamic events (e.g. split, merge) arise without any changes in the relevant regions of the network, but solely due to changes in hyperparameter values between time steps. To select the hyperparameters, we perform a small grid search over the OSLOM resolution and threshold parameters for each time step graph. We record the parameter values that maximise the combined fitness of the 10 largest communities. The pair of resolution and threshold values that occur most frequently across all time steps is then used for community detection across the entire dynamic network.

Hyperparameters are chosen to maximise the `fitness' (i.e., the proportion of edges with at least one end point in the community, that have both endpoints in the community \cite{lee2010detecting}) of the largest communities across all time steps. This focuses the analysis on the dominant, established research areas which tend to be largest, while still allowing new communities representing emerging topics to form over time. 
The choice of the specific number of communities (10 in this case) can be tailored based on the analysis goals. For example, given the nature of our dataset, many nodes may exist in early time steps that are not yet relevant to the rest of the literature. These papers typically represent unrealised applications of methods in the field and present as small fractured or isolated components in the citation network. Thus, we de-emphasise these smaller, isolated communities when evaluating fitness, preferring to prioritise the detection of the larger, more established topics. For rapidly evolving fields, focusing on the largest communities allows us to capture the major research areas while allowing smaller emerging areas to form.

\subsubsection{Labelling Step Communities}
\label{sec:labelling_communities}
Following \cite{jung2013analyzing,tan2022tracking}, we use the title and abstract text of the papers in a community to annotate the community's topic. Specifically, we combine the terms from all the papers in a community into a bag-of-words vector. We then annotate each community using the top-$n$ terms according to the Term Frequency inverse Community Frequency (TF-ICF). The ICF terms used to adjust the term frequencies are calculated per year. 
In addition we consider the category term descriptor (CTD) \cite{how2004empirical}. We treat each community as a category and calculate CTD based on: 
\begin{equation}
\label{eq:ctd}
\begin{split}
    CTD(t_k,C_i) = TF(t_k,C_i) \cdot IDF(t_k,C_i) \cdot ICF(t_k) \\ \text{where }\quad ICF(t_k) =  \log(\frac{\mid \mathbb{C} \mid}{CF(t_k)}),\quad  IDF(t_k,C_i) = \log(\frac{\mid c_i \mid}{DF(t_k,C_i)})
\end{split}
\end{equation}
Here $C$ denotes the set of communities, $CF(t_k)$ is the community frequency of term $t_k$, and $DF(t_k,C_i)$ is the document frequency for term $t_k$ in community $C_i$.
 
\subsubsection{Characterising Step Communities}
\label{sec:characterising_step_communities}

To characterise the step communities, we measure two key properties:~topic coherence and citation density.
Firstly, \textit{topic coherence} relates to the concept of the topic disparity \cite{kim2022quantifying} for a set of articles and more broadly to the notion of coherence in topic modelling \cite{roder2015exploring}. Initially, we learn a topic embedding for each article by passing the article title and abstract through a transformer-based language model trained on scientific articles (\textit{SciBERT}), and taking the final hidden state of the [CLS] token. We then compute the topic centroid for the community by taking the mean of all of the topic embeddings. Finally, we calculate the topic coherence of the community as the average similarity between each article's topic embedding and the community topic centroid. 
Our second measure, \textit{citation density}, refers to the network density of the citation subgraph described by a given community. This is measured as the number of edges (or citations) in the subgraph, divided by the number of possible connections. 

\subsubsection{Measuring Knowledge Transfer}
\label{sec:community_interaction_network}
In addition to characterising individual communities, at each time step we construct a \textit{community interaction network} $I_t = (\mathbb{C}t, J_t)$ to model knowledge transfer between the identified research areas. 
In the interaction network at time step $t$, the set of nodes corresponds to the step communities discovered at time $t$ and the set of edges correspond to the citations between them. Formally, $I_t$ has the set of nodes $\mathbb{C}_t$, and the set of weighted edges $J_t$ such that $J_t = \{(C_i^t,C_j^t,p_{ij}^t)\mid(v,u) \in E_t, v \in C_i^t, u \in C_j^t\}$. Here $p_{ij}^t$ is the probability of a interaction/citation between papers in $C_i^t$ and $C_j^t$, given by: 
\[p_{ij}^t = \frac{\mid \{(u,v) \in E_t \mid u \in C^t_i, v \in C^t_j \} \mid}{\mid C_i^t \mid \cdot \mid C_j^t \mid}\]
While the strength of the connections in the community interaction network network reveal the extent of the knowledge transfer between two research areas, we can also consider standard network measures to summarise the nature of knowledge transfer for any given research area. Specifically, we rely on two primary perspectives to summarise knowledge transfer for a research area. 
First, \textit{network centrality measures} reveal the extent to which topics are involved for the transfer of knowledge in the network. The unweighted degree centrality of a community in the interaction network reports the number of communities with which that topic shares knowledge, while the betweenness centrality measures the importance of that node in facilitating knowledge transfer across the network. Second, \textit{network proximity measures} help us to understand citation behaviours between two specified topics or communities. The connection strength between two communities in the interaction network is considered as the `first-order' network proximity. According to the above definition of the interaction network, the connection strength or edge weight between two communities is the probability of a citation between a paper in each community. `Second-order' network proximity of two communities is a measure of the similarity of the neighbourhoods of the communities. Thus, two communities will have high second-order proximity if they have similar citation behaviours with the other topics in the network. We measure second order proximity as the cosine similarity between the communities interaction probabilities across the network, i.e., between their respective columns of the weighted adjacency matrix describing the community interaction network. 

\subsection{Dynamic Communities}
The previous sections detailed how we identify step communities in individual time step graphs. A key goal is to track the evolution of these communities over time. This section outlines our approach for constructing dynamic community life-cycles by linking step communities across time steps. This allows us to analyse how research areas emerge, grow, merge, split or dissipate as the field evolves.

\subsubsection{Finding and Tracking Dynamic Communities}
\label{sec:tracking_communities}
We follow the method proposed in \cite{greene2010tracking} to track the life-cycles of communities in a cumulative citation network. In this framework, the step communities in the time step graphs represent observations of dynamic communities at a given time point (year). If we denote the set of step communities identified by OSLOM at time $t$ as $\mathbb{C}_t = \{C_1^t, C_2^t, .... C_n^t\}$, a dynamic community can be then represented by a chronology of step communities, for example $\mathbb{D}_1 = \{C_1^1, C_2^2, C_1^3\}$. At the first time point $t_0$, dynamic communities are formed using a one-to-one with the step communities $\mathbb{C}_0$. Subsequent step communities are added to the dynamic communities using a heuristic, many-to-many mapping. The most recent step community in a dynamic community is called its `front'. At a given time step, comparing the step communities with all of the dynamic community fronts can lead to a number of possible events: 
\begin{enumerate}
    \item If a step community does not match with any of the dynamic community fronts, it is added as a new dynamic community with a single step. This is known as community \textit{birth}.
    \item If a step community matches with a single front, that step community is added to that dynamic community timeline and becomes the new front. 
    \item If two or more step communities match with the same front, then new identical dynamic communities are formed and one of the matching step communities is added to each to act as it's front. This is known as community \textit{splitting}.
    \item If a step community matches with multiple fronts, then it is added to each of them. This is known as community \textit{merging}.
    \item If a dynamic community front does not match with any of the step communities then the front is not updated. The front may match with step communities at subsequent time steps, thus allowing for intermittent community structures to be found. If the front does not match with any of the step communities in any of the subsequent time steps then this is known as community \textit{death}. 
\end{enumerate}
To match community fronts with step communities, we follow the strategy proposed by \cite{greene2010tracking}. Given a step community $C_i^t$ and dynamic community front $F_j$, we compute the similarity between $C_i^t$, $F_j$ as:
\[sim(C_i^t,F_j) = \frac{\mid C_i^t \cap F_j\mid}{\mid C_i^t \cup F_j\mid}\]
Using the above measure for similarity, we match step communities to front if the similarity exceeds a matching threshold $\theta \in [0,1]$.

\subsubsection{Characterising Dynamic Communities}
\label{sec:characterising_dynamic_communities}

To analyse the properties and evolution of dynamic communities tracked across time steps, we propose a set of six metrics. The first metric is the community lifespan, measured as the number of time steps in which it is present in the dynamic network.
We then consider four metrics derived from time series data of the community's constituent step communities at each time point. The community size time series tracks the number of papers belonging to the step communities over time. The degree centrality and betweenness centrality time series measure how central the step communities are within the community interaction network in facilitating knowledge flow.
The final two metrics aim to quantify the stability and coherence of a dynamic community's research focus as it evolves, similar to existing approaches taken in NLP-based studies of research topic evolution \cite{song2014analyzing, taher2023embedding}. The content coherence metric compares the textual similarity of papers between consecutive step communities. Specifically, for a dynamic community $D = {C_1,\dots,C_i^{t-1}, C_j^t}$, we compute the average pairwise cosine similarity between the SciBERT topic embeddings of papers in $C_i^{t-1}$ and $C_j^t$ at each time step $t$. This yields a time series capturing how coherent the research topic remains.
Similarly, the membership stability metric tracks changes in the specific paper membership of the community over time. It is calculated as the Jaccard similarity between the paper sets of consecutive step communities $C_i^{t-1}$ and $C_j^t$ in the dynamic community $D$. \[sim(C_i^{t-1},C_j^t) = \frac{\mid C_j^t \cap C_i^{t-1}\mid}{\mid C_j^t \cup C_i^{t-1}\mid}\]
While metrics like size and centrality characterise the community's position and importance in the network, the content coherence and membership stability allow analysing the thematic evolution of the underlying research area. In most cases, we summarise the time series values using averages over the community's lifespan or specific time periods of interest.

\subsection{Research Questions}
\label{sec:research_questions}
Given the above described framework for discovering, interpreting and characterising research areas in some corpus, in this section we outline how our proposed methods can be used to answer our three key research questions from Section~\ref{sec:intro} relating to knowledge transfer in XAI research. 

\begin{enumerate}
\item \textit{To what extent do contemporary topics in the literature rely on foundational research in psychology, statistics and computer science?}
We answer this question in a number of steps: 
\begin{enumerate}
    \item Identify the foundational topics in the literature as \textit{long-lived} communities with \textit{coherent subject matter} that are consistently \textit{central} in the interaction networks.
    \item Identify contemporary topics as the communities present in the later periods of the dataset that are populated by the most recent papers. 
    \item Separate and compare those recent communities that cite the foundational topics from those that do not. 
\end{enumerate}

\item \textit{What are the research areas that are most isolated in terms of knowledge transfer within the literature?} We refer to these isolated research areas as `knowledge silos', and we identify them as the nodes with the lowest total interaction probability in the community interaction network.  

\item \textit{Is there evidence of knowledge gaps between otherwise related research areas?/What are the most significant knowledge gaps in the literature?}
Our approach to identifying knowledge gaps is outlined below: 
\begin{enumerate}
    \item Use a regression model to predict interaction probabilities in the community interaction network based on the research areas' content similarity and citation neighbourhood proximity.
    Specifically, we use a regression model with a gamma distribution (implemented in Python 3.9.7 using \textit{Scikit-learn} \cite{pedregosa2011sklearn}) to predict the interaction probabilities between all pairs of communities in the final time step graph. The independent variables are: 1) the average pairwise cosine similarity between the SciBERT embeddings of papers in each community (content similarity), and 2) the cosine similarity between the communities' connection probabilities in the interaction network (second-order network proximity capturing similarity of citation neighbourhoods). SciBERT embeddings are learned from the papers' title and text using a pre-trained transformer language model (SciBERT \cite{beltagy2019scibert}) provided by \textit{HuggingFace}.
    \item Analyse the residuals: Pairs of communities with large positive residuals from the model predictions are then identified as having knowledge gaps, since they demonstrate far less knowledge transfer than expected given their content relatedness and structural proximity in the network.
    \item Highlight research areas that exhibit multiple large positive residuals and examine these knowledge gaps. 
\end{enumerate}
\end{enumerate}

\subsection{Data}
\label{sec:data}

To demonstrate the utility of the methods described above we chose eXplainable Artificial Intelligence (XAI) research as a case study of a rapidly growing field of research that is highly multidisciplinary.
We seed our data collection process with the \textit{xai-scholar} dataset \cite{jacovi2023trends}, which was collected on December 31st 2022.
The \textit{xai-scholar} dataset was curated using a process of keyword-based search, manual expansion, citation expansion and keyword-based filtering to produce a dataset of XAI 5,119 papers \cite{jacovi2023trends}. 
For the purposes of our analysis, and to obtain a more complete view of the citation network,  it was necessary to expand the \textit{xai-scholar} dataset to contain `non-XAI' works that are related to XAI research. 
Many of the earliest works in the field of XAI predate the modern terminology, and do not self identify as XAI research. Further, any works that are heavily cited by these papers are deemed relevant to the field and necessary to include in the citation graph to gain an accurate understanding of knowledge transfer in the literature. 
As such, we extend the \textit{xai-scholar} citation network using metadata from the \textit{Semantic Scholar Academic Graph} API \cite{kinney2023scholar}, using a 1-hop citation expansion, and retain any papers that have a citation relationship (citing or cited by) with more than one paper in the core set of \textit{xai-scholar} papers. The resulting citation network contains 20,604 papers published between 1889 and 2023 and 306,668 citations. The data collection from the Semantic Scholar Academic Graph was completed in November 2023, and the metadata for the final set of papers considered in this work is available on GitHub \cite{data_code}. 

\section{Results}
\label{sec:results}

The presentation of our results is structured as follows. Firstly, Section \ref{sec:preliminary} presents an initial analysis to visualise the life-cycles of illustrative examples of dynamic research communities. In Section \ref{sec:knowledge_transfer} we describe the identification of foundational areas in the XAI literature and assess the extent to which contemporary XAI research topics build upon and integrate knowledge from these foundational areas. In Section \ref{sec:silos}, we identify potential knowledge silos -- isolated research areas exhibiting minimal knowledge transfer with the rest of the field.  Lastly, Section \ref{sec:gaps} models citation interactions between research areas to detect significant knowledge gaps where insufficient transfer occurs among otherwise related topics.

\subsection{Preliminary Results -- Community Life-cycles}
\label{sec:preliminary}

\begin{figure}
    \centering
    \includegraphics[width=\linewidth]{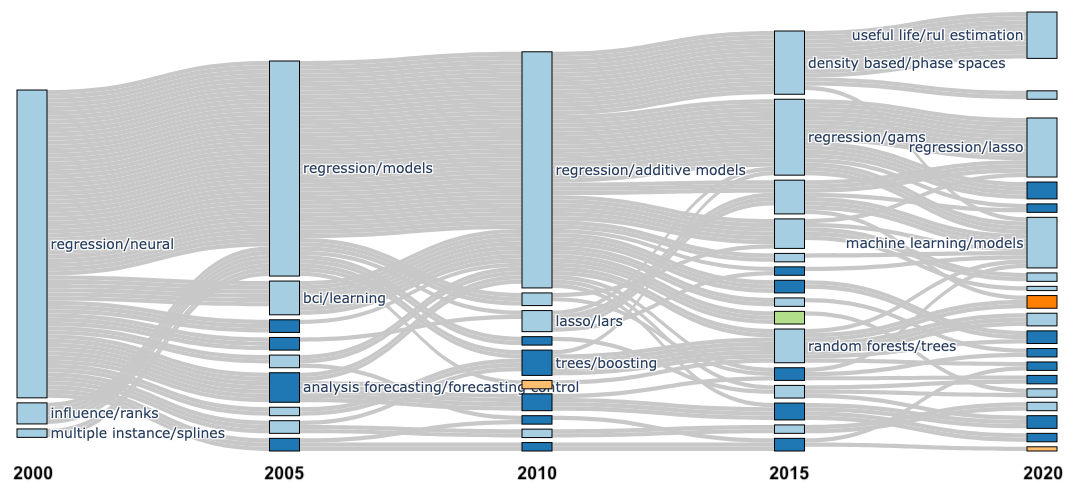}
    \caption{Flow diagram showing the life-cycle of a dynamic community pertaining to statistics research on regression models. Any papers present in the first or last realisation of the dynamic community are plotted. The nodes in the graph represent step communities and they are grouped by the time step in which they appear. The edges between nodes show the movement of papers between step communities. The dynamic community dissolves after 2010}
    \label{fig:flow_regression}
\end{figure}

\begin{figure}
    \centering
    \includegraphics[width=\linewidth]{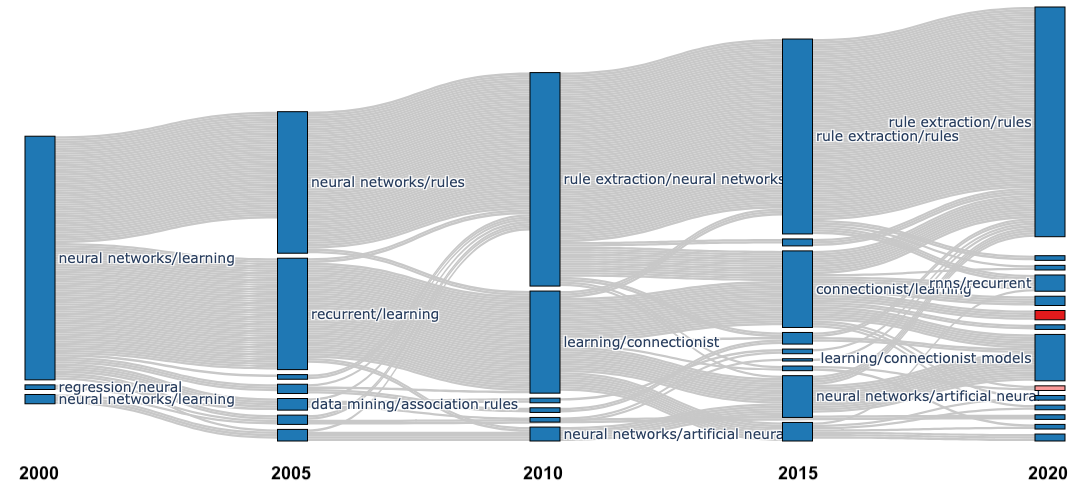}
    \caption{Flow diagram showing the life-cycle of a dynamic community of neural networks research that splits before 2005. The research area that focuses on rule extraction remains present in 2020. Any papers present in the first or last realisation of the dynamic community are plotted. The nodes in the graph represent step communities and they are grouped by the time step in which they appear. The edges between nodes show the movement of papers between step communities.}
    \label{fig:flow_neural_rule}
\end{figure}

In this analysis, since we investigate knowledge transfer in the area of XAI-related research, we consider communities discovered at the lowest level of the OSLOM hierarchy. 
To demonstrate our approach and to provide context for our definition and discovery of \textit{research areas}, we provide two examples of community life-cycles in the flow diagrams in Figures \ref{fig:flow_regression} and \ref{fig:flow_neural_rule}. In total we identify life-cycles for 435 dynamic communities or research areas. Of these areas, 163 dissolve before the final time step in 2023. Research area sizes range from 3 to 320 papers, with a median size of 50. We choose the two examples in Figures \ref{fig:flow_regression} and \ref{fig:flow_neural_rule}, as they represent long-lived communities with different characteristics. Moreover, both of these example research areas are relevant to discussion in Section \ref{sec:knowledge_transfer} as they are identified as foundation topics in the XAI literature (See Section \ref{sec:knowledge_transfer} for details). The flow diagram presented in Figure \ref{fig:flow_regression} shows the life-cycle of a research area. In particular, the figures shows how the research area relating to regression models `dissolves', as its constituent papers are cited by work in many other topics. Dynamic community `death' -- as it is described in the social network analysis literature -- has some specific caveats in the context of citation network analysis. Specifically, as nodes and connections are not removed from the network, community `death' only occurs in the form presented in Figure \ref{fig:flow_regression}, where a community dissolves as the connections between community's members are surpassed by connections to external works. Crucially, we note that the community dissolution shown here is specific to perspective of our dataset, i.e., of XAI-related research. 

For comparison, we include the life-cycle of another, more stable, research area in Figure \ref{fig:flow_neural_rule}. In this example, we can see that the community of neural network research splits some time before 2005 into works on `rule extraction' and `recurrent' `connectionist architectures'. The `rule extraction' community continues to grow steadily after this point, while the remaining connectionist computing community dissolves into various applications and sub-fields. 

\subsection{Knowledge Transfer from Foundational Research Areas}
\label{sec:knowledge_transfer}
\begin{figure}[!t]
    \centering
    \includegraphics[width=\linewidth]{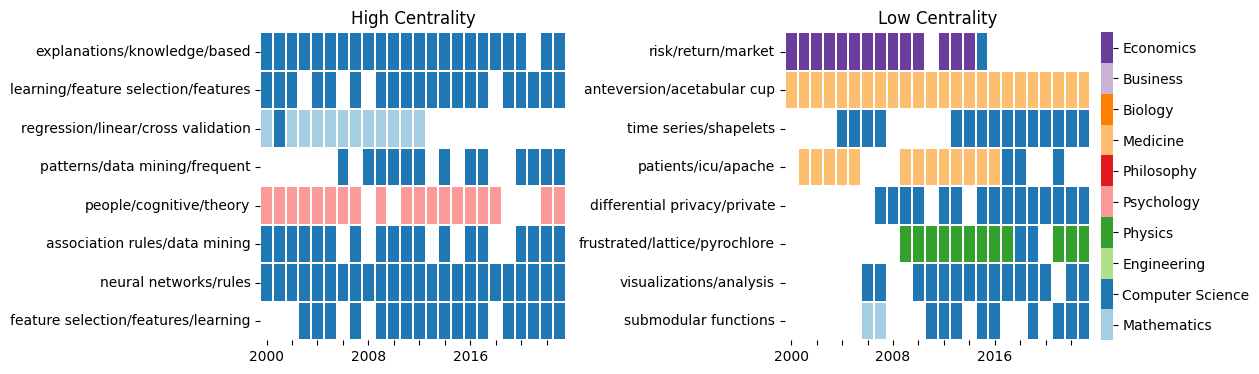}
    \caption{Dynamic community life-cycles of the communities with the highest betweenness centrality (left) and lowest betweenness centrality (right) in the period 2000--2010. Each row represents the life-cycle of a dynamic community and each cell in the row is populated if that dynamic community appears in the network as a step community in the corresponding time step. Each cell is coloured to show the most common ASJC category among the papers in the step community.}
    \label{fig:high_low_betweenness}
\end{figure}

We now assess the extent to which contemporary research areas in the field of XAI rely on the theoretical and methodological foundations of the field. 
Firstly, we identify foundational areas in the literature as long-lived dynamic communities, which are important to the knowledge transfer in the field in the earlier portions of the dataset. Figure \ref{fig:high_low_betweenness} shows a simplified view of the life-cycles of the dynamic communities with the highest (left) and lowest (right) betweenness centrality scores, as measured in the community interaction network in the period 2000--2010. We rely on centrality in the community interaction network (see Section \ref{sec:community_interaction_network}) to reveal the relative importance of the research area for knowledge transfer in the network over time. For example, all dynamic communities shown in Figure \ref{fig:high_low_betweenness} are relatively long-lived and stable. However, topics in Economics, Physics and Medicine, remain isolated in the early portion of the dataset, as they represent later applications of XAI research. Conversely, research in the areas of Computer Science, Statistics and Psychology, have high betweenness centrality in the early period, as they engage in knowledge transfer consistently throughout the development of the field. Thus we conclude these research areas to represent the methodological and theoretical foundations of XAI literature.  

\begin{figure}[!t]
    \centering
    \includegraphics[width=\linewidth]{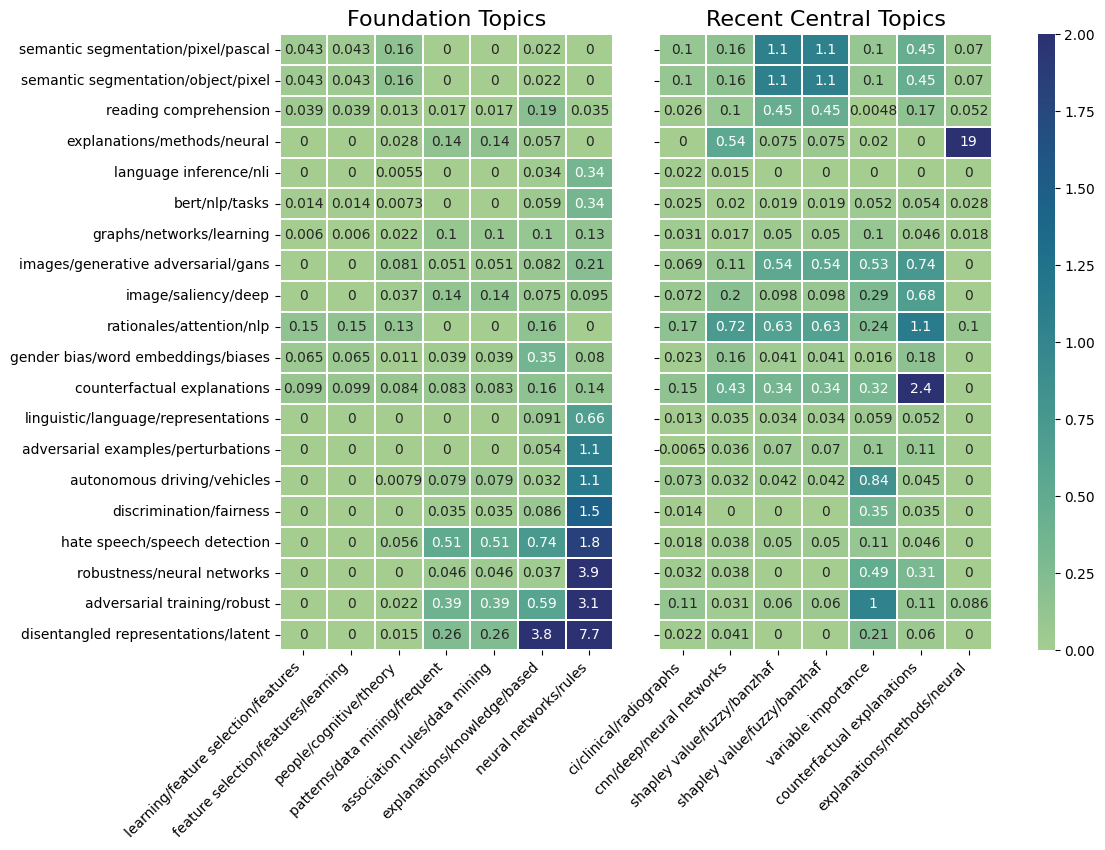}
    \caption{The percentage of total possible interactions (citations) between contemporary research ares in XAI literature and the foundation areas. For comparison, some recent central topics in XAI are included on the right. For readability, the interaction probabilities are scaled to percentages. For example, a score of 2\% between two research areas indicates that 20\% of papers in research area A cite 10\% of papers in research area B.}
    \label{fig:topic_interactions}
\end{figure}

Secondly, we assess the extent to which contemporary research areas in XAI leverage knowledge and methodologies from foundational research in the field. We identify the contemporary research areas in the literature as dynamic communities where the average age of the papers is six years or fewer (i.e., published since 2017). We focus our analysis on large communities (greater than 50 papers), with a content stability score above the mean measured across the dataset. Thus, we consider large communities of recent research papers with coherent content/topic as the clearest representations of the contemporary research in the literature. We assess how these research areas rely on foundational research using interactions in the community interaction network measured at the most recent time step (2023). Figure \ref{fig:topic_interactions} reports the interaction (citation) probabilities between the largest contemporary research areas in the field and the foundations identified previously. For comparison, we include equivalent scores measured between the contemporary research topics and more recent central topics, corresponding to the dynamic communities with the highest betweenness centrality during 2020--2023.

\begin{figure}[!t]
    \centering
    \includegraphics[width=\linewidth]{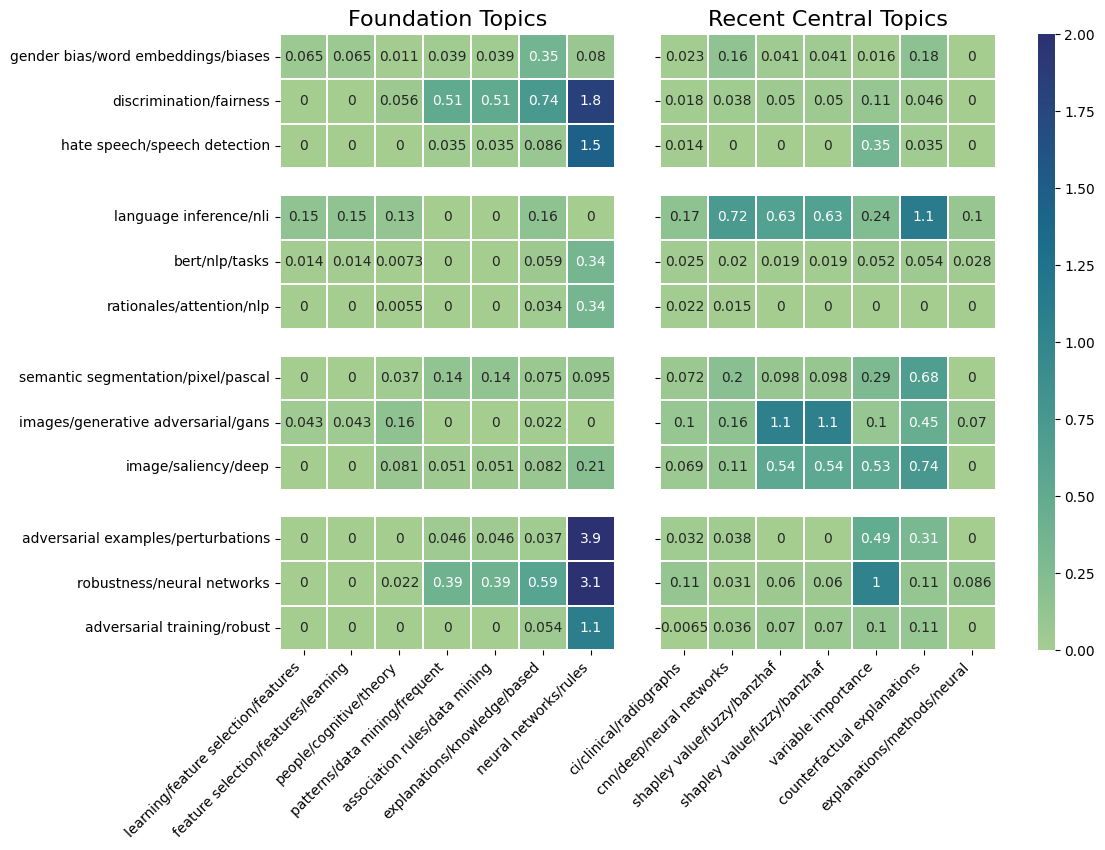}
    \caption{The percentage of total possible interactions (citations) between contemporary research areas in XAI literature and the foundation areas. For comparison, some recent central topics in XAI are included on the right. For readability, the interaction probabilities are scaled to percentages. For example, if 20\% of papers in research area A cite 10\% of papers in research area B, the resulting interactions score would be 2\%.}
    \label{fig:topic_interactions_sub}
\end{figure}

In Figure \ref{fig:topic_interactions_sub}, we highlight four subsets of the contemporary research areas which have similar content to compare how they interact with the established literature. In particular, we group research areas into four research topics relating to (i) \textit{fairness}, (ii) \textit{natural language processing}, (iii) \textit{computer vision}, and (iv) \textit{adversarial machine learning}. Thus, we reveal some patterns in knowledge transfer by topic. For example, research areas within the `\textit{computer vision}' group exhibit similar knowledge transfer behaviour as they rely more on knowledge from the more recently central topics, than they do from the historically central or `\textit{foundational}' topics. Conversely, research areas in the `\textit{fairness}' group demonstrate different knowledge transfer patterns despite their related content. In particular, the two research areas labelled `\textit{hate speech}' and `\textit{gender bias}' make fewer citations to methodological foundations of XAI research than the third research area (labelled `discrimination/fairness'). 

\subsection{Knowledge Silos}
\label{sec:silos}

We now illustrate the identification of knowledge silos, corresponding to isolated research areas in the community interaction network. Table \ref{tab:silos} reports summary statistics for the five research areas with the lowest sum across all interaction probabilities in the final (most recent) snapshot of the community interaction network (snapshot 2023). During this process We exclude small research areas made up of 10 or fewer papers. 

\begin{table}[!t]
    \centering   
\resizebox{\textwidth}{!}{
\begin{tabular}{llrrr}
\toprule
 & \textbf{Label} & \textbf{Size} & \textbf{Degree} & \textbf{Density} \\
\midrule
1 & learning/feature selection/features/algorithm/classification & 255 & 188 & 0.03 \\
2 & feature interactions/ctr prediction/click/fms/factorization machines & 33 & 62 & 0.37 \\
3 & soil/water/n2o emissions/hydrological/nitrous oxide & 107 & 128 & 0.08 \\
4 & covid/coronavirus/pneumonia/patients/ray images & 296 & 165 & 0.05 \\
5 & probast/prediction model/patients/disease/survival & 65 & 67 & 0.05 \\
\bottomrule
\end{tabular}}
    \caption{Knowledge silos: The top five most isolated research areas in the community interaction network in 2023. }
    \label{tab:silos}
\end{table}

XAI applications in \textit{Environmental Sciences/Atmospheric Chemistry and Physics} and \textit{COVID-19 diagnosis from chest CT images} are identified as two of the most isolated research areas in the corpus, with respect to knowledge transfer. Each of these research areas represent intuitive applications of XAI and machine learning research to important real-world challenges. However, the isolated position of these areas in the community networks is problematic. In fact, recent studies have highlighted limitations to the utility of the each of these applications in their respective domains \cite{roberts2021common,silva2024limitations}. Specifically, Silva and Keller show that dependencies and strong correlations between features lead to model explanations that are inconsistent with process-level understanding \cite{silva2024limitations}. Similarly, in their review of computer vision solutions for COVID-19 detection, Roberts at al.~find that none of the solutions are clinically viable, due to important methodological flaws in the machine learning applications \cite{roberts2021common}. 
These cases represent examples of some of the practical issues with isolated research areas and limited knowledge transfer in interdisciplinary applications. They highlight that unrealised knowledge integration, either from the domain of the application (as in \cite{silva2024limitations}), or from the domain of the methods (as in \cite{roberts2021common}), can lead to poorer outcomes in terms of the utility of the applications.

\subsection{Knowledge Gaps}
\label{sec:gaps}

To identify potential knowledge gaps between research areas in the XAI literature, we modelled the expected knowledge transfer between communities based on their content similarity and proximity in the community interaction network. Pairs of communities exhibiting substantially lower interaction (citation) rates than predicted were flagged as having significant knowledge gaps.
The model essentially captures the relationship between research areas' content similarity, the similarity between their knowledge sources and knowledge outputs, and their observed knowledge transfer behaviour. By analysing the residuals, we identify those pairs of communities that demonstrate less knowledge transfer than we would expect based on their content and related work. 
This approach allows detecting gaps in knowledge flow that may arise due to disciplinary boundaries, insular reading and citation patterns, or lack of awareness of complementary work.

\begin{table}[!t]
    \centering   
\begin{tabular}{ll}
\toprule
 & \textbf{submodularity/contrastive explanations} \\
\midrule
1 & deep/learning/neural networks/classification/convolutional neural \\
2 & machine learning/variable importance/feature/random \\
3 & conversational/dialog/conversation/chatbot/language \\
4 & segmentation/brain/tumor/medical image/learning \\
5 & object/scene/images/visual/recognition \\
\midrule
 & \textbf{visualization/visual analytics/learning} \\
\midrule
1 & adversarial training/adversarial examples/attacks/robust/perturbations \\
2 & fine grained/grained recognition/cub/parts/image \\
3 & machine translation/attention/nmt/neural machine/sequence \\
4 & machine learning/comprehensibility/classification/support vector/methods \\
5 & type fuzzy/fuzzy logic/fpi/tsk fuzzy/deep \\
\midrule
 & \textbf{counterfactual explanations/recourse} \\
\midrule
1 & segmentation/brain/tumor/medical image/learning \\
2 & deep/learning/neural networks/classification/convolutional neural \\
3 & object/scene/images/visual/recognition \\
4 & object detection/cnn/learning/coco/detectors \\
5 & conversational/dialog/conversation/chatbot/language \\
\midrule
 & \textbf{segmentation/brain/tumor/medical image} \\
\midrule
1 & counterfactual explanations/recourse/counterfactuals/learning \\
2 & submodularity/contrastive explanations/submodular functions/local contrastive/greedy \\
3 & clinical/medical/ai/deep learning/radiographs \\
4 & explanations/idealization/scientific/mental models/understanding \\
5 & machine learning/variable importance/feature/random \\
\bottomrule
\end{tabular}
    \caption{Knowledge gaps: The four research areas that most consistently demonstrate knowledge gaps. The research areas marked in bold are the four areas that have the greatest total residual score (indicating overestimated knowledge transfer) across all possible research areas. In the case of each area, we include the five research areas with which they demonstrate the most significant knowledge gaps.}
    \label{tab:gaps}
\end{table}

Table \ref{tab:gaps} highlights the four communities that most consistently demonstrate knowledge gaps (i.e., have the greatest total residual score across all possible research areas) and the research areas with which they have the most significant gaps.
We recognise that the knowledge gaps identified can be categorised into two groups. (i) Between methodological research areas and potential applications (e.g. between \textit{counterfactual explanations} and multiple research areas in \textit{computer vision} or between \textit{contrastive explanations} and research ares in \textit{natural language processing}). (ii) Between two applied research areas studying the same or similar topics (e.g. \textit{computer vision for medical images}).
In the case of (ii), we highlight one of the key benefits to our proposed methods for delineating \textit{research areas}. NLP based methods for recognising \textit{research topics} would (correctly) group these research areas into a single topic. When studying knowledge transfer, it is important to recognise that they are separate from one another, but pertain to the same or similar topic. Thus, we recognise the benefits of studying citation relations and article content in tandem. 

\section{Discussion}
This work introduces a novel network analysis framework to study the dynamics of knowledge transfer and integration in rapidly evolving, interdisciplinary research fields. By applying dynamic community detection techniques to citation networks, we can identify and track the emergence, evolution, and interactions of research areas or topics directly from the published literature. The key methodological contributions include:
\begin{enumerate}
    \item Providing a new perspective on dynamic community finding algorithms to facilitate their application to the unique context of citation networks, which exhibit cumulative growth and content-rich nodes (papers).
    \item Developing methods to characterise the properties of identified dynamic communities over time, such as content coherence, and knowledge transfer centrality. These methods begin to bridge existing gaps between the citation network-based approaches to mapping research areas and the traditional NLP-based methods. 
    \item Analysing the interactions between dynamic communities using our proposed \textit{community interaction network} to reveal patterns of knowledge transfer, isolate potential knowledge silos, and detect significant knowledge gaps.
    
\end{enumerate}

\noindent We demonstrated the utility of our approach through a case study on eXplainable Artificial Intelligence (XAI) -- an emerging, highly interdisciplinary field synthesising concepts from machine learning, psychology, philosophy and other domains. The key findings include:
\begin{enumerate}
    \item Foundational areas, such as statistics, cognitive science, and interpretable machine learning, acted as important knowledge sources during the formation of the field of XAI. However, knowledge transfer between these areas and the contemporary topics in XAI research is limited and the extent of knowledge transfer varies across different contemporary research topics.
    \item Certain research areas like applications in COVID-19 diagnosis and environmental science exhibit characteristics of knowledge silos, as they remain isolated from the knowledge transfer exhibited between the rest of the XAI-related research areas.   
    Limitations to the utility of these applications have been highlighted by recent studies \cite{roberts2021common,silva2024limitations}.
    \item Notable knowledge gaps were identified throughout the literature, falling under two themes broad themes: (i) Between methodological research areas and potential applications (e.g. between \textit{counterfactual explanations} and multiple research areas in \textit{computer vision} or between \textit{contrastive explanations} and research ares in \textit{natural language processing}). (ii) Between two applied research areas studying the same or similar topics (e.g. \textit{computer vision for medical images}).
\end{enumerate}

\noindent By mapping the flows, interdisciplinary integration, and boundaries of this evolving field, our analysis can inform strategies to promote cross-pollination, bridge disciplinary divides, and synthesise disparate ideas to drive innovation in XAI research and applications.
More broadly, this work provides a data-driven framework to study the evolution of knowledge ecosystems and the dynamics of interdisciplinary integration directly from the published literature. The methodological contributions have applications spanning "science of science" studies, literature review and analysis, interdisciplinary research planning, and science policy and funding decisions. As scientific fields become increasingly specialised yet coupled, tools to understand and facilitate knowledge transfer across disciplines will become ever more critical. This work establishes a novel network analysis-based approach towards that important goal.

\bibliographystyle{acm}
\bibliography{references}
\section*{Acknowledgements}
This research was conducted with the financial support of Science Foundation Ireland under Grant Number 12/RC/2289\_P2 at the Insight SFI Research Centre for Data Analytics. 
\end{document}